\documentclass{article}

\usepackage{arxiv}

\usepackage[utf8]{inputenc} % allow utf-8 input
\usepackage[T1]{fontenc}    % use 8-bit T1 fonts
\usepackage{hyperref}       % hyperlinks
\usepackage{url}            % simple URL typesetting
\usepackage{booktabs}       % professional-quality tables
\usepackage{amsfonts}       % blackboard math symbols
\usepackage{nicefrac}       % compact symbols for 1/2, etc.
\usepackage{lipsum}
\usepackage{graphicx}
\graphicspath{ {./images/} }

\usepackage{tabularx}
\usepackage{float} 
 
\usepackage[acronym]{glossaries}
\makeglossaries

\newacronym{ndt}{NDT}{Network Digital Twin}
\newacronym{npt}{NPT}{Network Physical Twin}
\newacronym{dt}{DT}{Digital Twin}
\newacronym{ue}{UE}{User Equipment}
\newacronym{sci}{SCI}{State Consistency Index}
\newacronym{mmtc}{mMTC}{massive Machine Type Communication}
\newacronym{embb}{eMBB}{Enhanced Mobile Broadband}
\newacronym{urllc}{URLLC}{Ultra-Reliable Low Latency Communication}
\newacronym{mec}{MEC}{Multi-Access Edge Computing}
\newacronym{ran}{RAN}{Radio Access Network}
\newacronym{sba}{SBA}{Service Based Architecture}
\newacronym{upf}{UPF}{User Plane Function}
\newacronym{qci}{QCI}{Quality of Service Class Identifiers}
\newacronym{gnb}{gNB}{gNodeB}
\newacronym{ims}{IMS}{IP Multimedia Subsystem}
\newacronym{snmp}{SNMP}{Simple Management Protocol}
\newacronym{ml}{ML}{Machine Learning}
\newacronym{ai}{AI}{Artificial Intelligence}
\newacronym{nn}{NN}{Neural Networks}
\newacronym{sdn}{SDN}{Software Defined Network}
\newacronym{nfvi}{NFVI}{Network Functions Virtualization Infrastructure}

\title{A Network Digital Twin of a 5G Private Network: Designing a Proof-of-Concept from Theory to Practice}

\author{
 Cristina Emilia Costa \\
  Consorzio Nazionale Interuniversitario per le Telecomunicazioni (CNIT) \\
  Smart and Secure Networks (S2N) Laboratory\\
  Genova, Italy \\
  \texttt{ccosta@cnit.it} \\
   \And
 Tatenda Horiro Zhou \\
  Dept of Information Engineering and Computer Science\\
  University of Trento\\
  Trento, Italy \\
  \texttt{tatendazho@gmail.com} \\
  \And
 Fabrizio Granelli \\
  Consorzio Nazionale Interuniversitario per le Telecomunicazioni (CNIT), \\
  and Dept of Information Engineering and Computer Science\\
  University of Trento\\
  Trento, Italy \\
  \texttt{fabrizio.granelli@unitn.it} \\
}

\begin{document}
\maketitle
\begin{abstract}
Network Digital Twins represent a key technology in future networks, expected to provide the capability to perform accurate analysis and predictions about the behaviour of 6G mobile networks. However, despite the availability of several theoretical works on the subject, still very few examples of actual implementations of Network Digital Twin are available. This paper provides a detailed description about the characteristics of Network Digital Twin and provides a practical example about real deployment of the technology. The considered network infrastructure is a real 5G private network running in a lab. The Network Digital Twin is built based on open source network emulation software and is available to the community as open source. Measurements on both the physical infrastructure and the related Digital Twin demonstrate a high accuracy in reproducing the state and behavior of the actual 5G system.
\end{abstract}


\section{Introduction and Literature Survey}

The Digital Twin concept have been adopted various fields, such as Industry 4.0, Smart Cities, etc. By applying Digital Twin concepts and techniques to the networks it is possible to create a \gls{ndt}, that becomes a virtual replica of real  network infrastructure (e.g. through emulation).  A \acrfull{ndt} can therefore be considered an  advanced platform for network emulation. It can serve as a tool for many network management operations such as scenario planning, impact analysis, and change management. However, with respect to conventional network simulation and emulation tools, the \gls{ndt} relays on real-time interaction and data exchange for keeping  \gls{ndt} with the \gls{npt} interactive in sync. Data between twins are constantly mapped and updates, thus allowing a data-driven approach to establish closed-loop   network automation \cite{sun2022introduction}. 
 
The Digital and Physical twins together with the data communication between them constitute a system, the \textbf{Network Twinning System}. The continuous interaction between the \gls{ndt} and the \gls{npt} is what differentiates the concept of network digital twin from a digital model or a simulation tool. Thanks to the interaction, the implemented model becomes dynamic and able to adapt to network changes. Network data in the loop can be integrated with external data, e.g. mobility and environmental data, that can be used for optimization and prediction of the usage of network, increasing its adaptability also to events external to the network.

\begin{figure}
    \centering
    \includegraphics[width=1\linewidth]{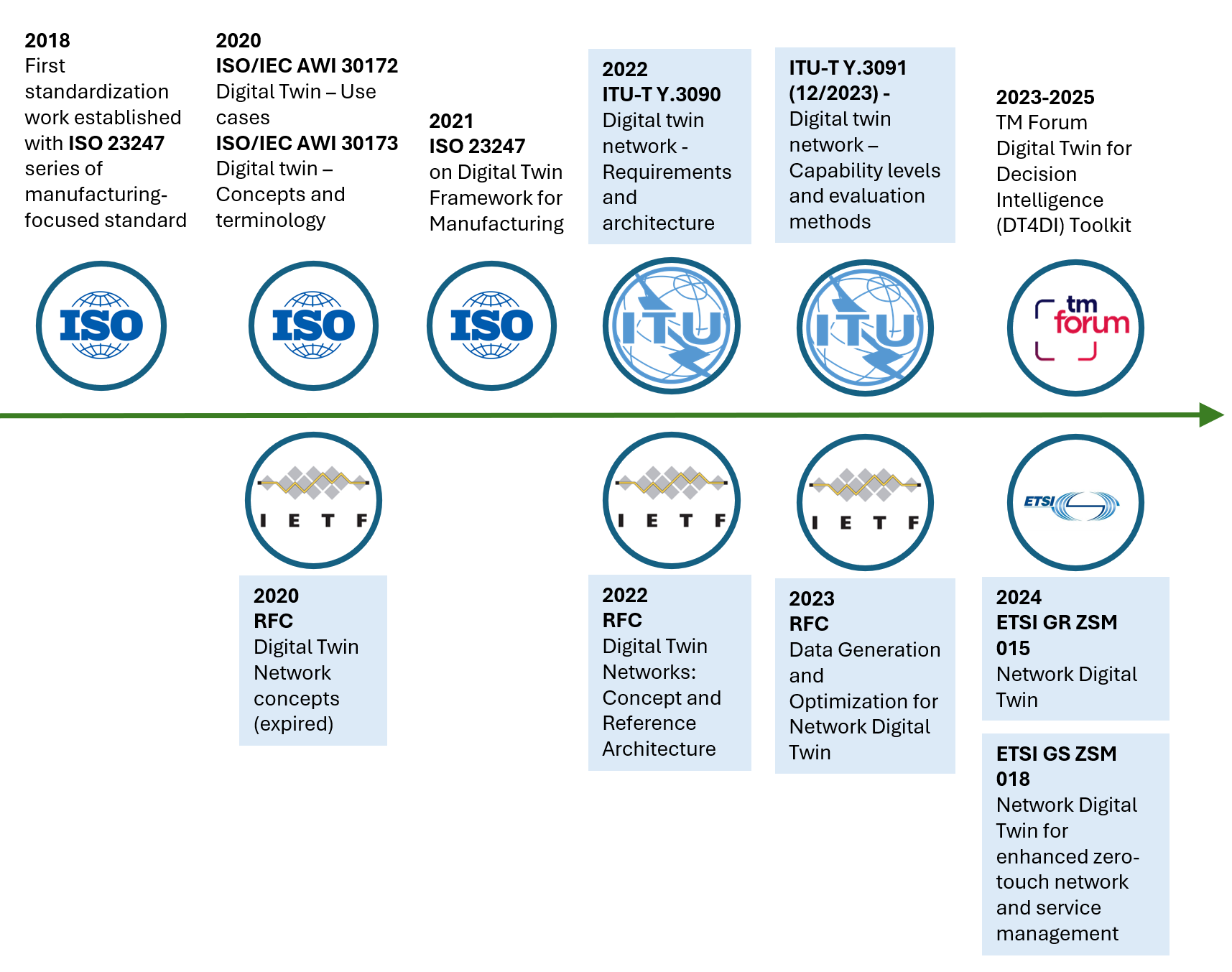}
    \caption{Network Digital Twin in the standardization process}
    \label{fig:stad-process}
\end{figure}

Digital Twins started to gain attention from the standardization bodies with the establishment of the ISO 23247 (\textit{Automation systems and integration — Digital Twin framework for manufacturing}) series in 2018 (Fig.\ref{fig:stad-process}), introducing a generic framework that can be specialized to different manufacturing processes scenarios (e.g., discrete, batch, or continuous) \cite{shao2023analysis}. Even if the work under ISO 23247 is focused on manufactured focused, it introduces views on general principles and architecture that are valuable for the study of all Digital Twins. 

In particular, in ISO 23247 Part 2 includes a Reference Architecture for a Digital Twin from domain point of view. In this architecture, the Digital Twin encompasses four domains:

\begin{itemize}
    \item Observable domain: placed outside from the Digital Twin, it is composed by the observable elements of the Physical Twin, provides a context for the digital twin development and interacts with the data collection and device control domain.
    \item Data collection and device control domain: linking the elements of the Physical Twin to their digital twins counterparts for synchronization by collecting data from the sensors and probes, and controlling and actuating on the Physical Twin.
    \item Core domain: responsible for overall operation and management of a Digital Twin. It hosts applications and services such as data analytics, simulation, and optimization to enable provisioning, monitoring, modeling, and synchronization. It is also provides a northbound interface towards the users of the digital twin and other digital twins.
    \item User domain: responsible for interaction of the users with the digital twins. A user can be a human, a device, an application or a system that uses applications and services provided by the digital twins.
\end{itemize}

The smooth communication between  devices, digital twins, and user systems \cite{iso-23247.4} is instead addressed by ISO 23247-4 "\textit{Information exchange}." It points its attention on the role of real-time data exchange, system compatibility, and layered architecture in bridging physical and digital operations effectively. Table \ref{tab:iso-23247-4} summaries some possible challenges in this information exchange.

\begin{table}[H] 
%\small % Change table font size
\caption{Challenges in information exchange between Digital and Physical Twins.\label{tab:iso-23247-4}}
%\isPreprints{\centering}{} % Only used for preprints
\begin{tabularx}{\textwidth}{XXXX}
\toprule
\textbf{Integration Challenge}	& \textbf{Impact on the Digital Twining System}	& \textbf{Possible approaches}\\
\midrule
Data Format Incompatibility	 & Disrupted Information Flow between Digital and Physical Twins or between Digital Twins &	Standardized data formats for information exchange \\
Network Connectivity Issues	& System Performance Problems, Synchronization failures & Network monitoring and strict maintenance plans. \\
Legacy System Integration	& Communication Barriers due to lock-in, interfaces that are not open	& Standard interfaces or custom adapters and middleware for seamless compatibility. \\
\bottomrule
\end{tabularx}
\end{table}

\begin{figure}[ht]
    \centering
    \includegraphics[width=1\linewidth]{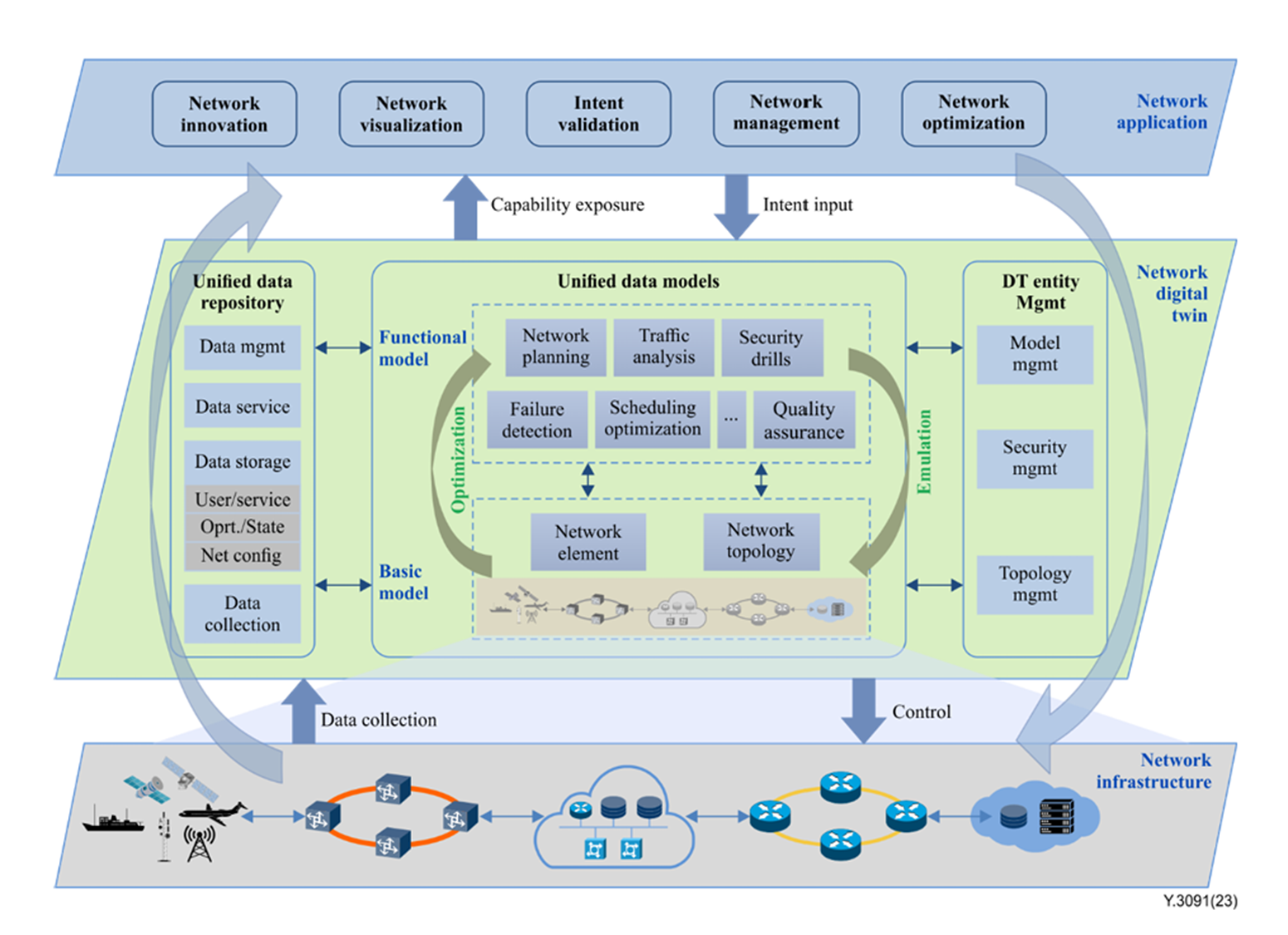}
    \caption{ITU Y.3090 \gls{ndt} Reference architecture (2022)}
    \label{fig:itu-Y.3090_ref-architecture}
\end{figure}

In 2020 the first the IETF Internet draft specifically focusing on the  \gls{ndt} domain was published. This was followed by an IETF Internet draft defining  a reference architecture for the \gls{ndt} \cite{draft-irtf-nmrg-network-digital-twin-arch}. Also ITU-T, in its specification Y.3090 defines requirements a \gls{ndt} and a reference architecture (Fig.\ref{fig:itu-Y.3090_ref-architecture}) aligned with the four domains defined in ISO 23247. Indeed we can relate the two as follows:

\begin{itemize}
    \item \textit{Observable domain} --> Network Infrastructure, the network physical twin
    \item \textit{Data collection and device control domain} --> Data collection from the network elements and control
    \item \textit{Core domain} --> Network digital twin composed by the unified data repository, unified data models and digital twin entity management
    \item \textit{User domain} --> Network application, including network innovation, network virtualization, intent validation, network management and network optimization.
\end{itemize}

The IETF Internet Draft on \textit{Data Generation and Optimization for Network Digital Twin Performance Modeling} that addresses how to improve the learning models in the \gls{ndt} from the data perspective. Indeed, AI models, like \gls{ml} and \gls{nn}, can be used in \gls{ndt} for various purposes such as performance modeling, optimization and prediction. The quality of network data sources can be very heterogeneous, being sources of varying nature. Many times, the collected data cannot directly serve for training \gls{ndt} models. Since the quality and volume of training data used directly affects the accuracy and generalization ability of the model, data generation and optimization methods can generate simulated network data to solve the problem of practical data shortage and select high- quality data from various data sources.

In \cite{draft-li-nmrg-dtn-data-generation-optimization} three sources of Network data are described:

\begin{enumerate}
    \item Data generated directly from production networks: this usually have high value, but the quantity, type, and accuracy are limited. In some scenarios collection of this type of data under various configurations may be unpractical;
    \item Data generated by network simulators, (e.g., NS-3 and OMNeT++): running a complex simulation scenario can be time and resource consuming but can fulfill, to a certain extend, the quantity, diversity, and accuracy requirements.  Differences between simulated data and practical data from production networks, should be taken into account and evaluated, depending on the application scenario;
    \item Data generated by Generative AI models (e.g., GPT and LLaMA): the simulated network data generated can, to a certain extend, fulfill the quantity and diversity requirements. However, the accuracy of such data may be is limited and often has gaps with practical data from production networks.
\end{enumerate}

\begin{figure}
    \centering
    \includegraphics[width=0.75\linewidth]{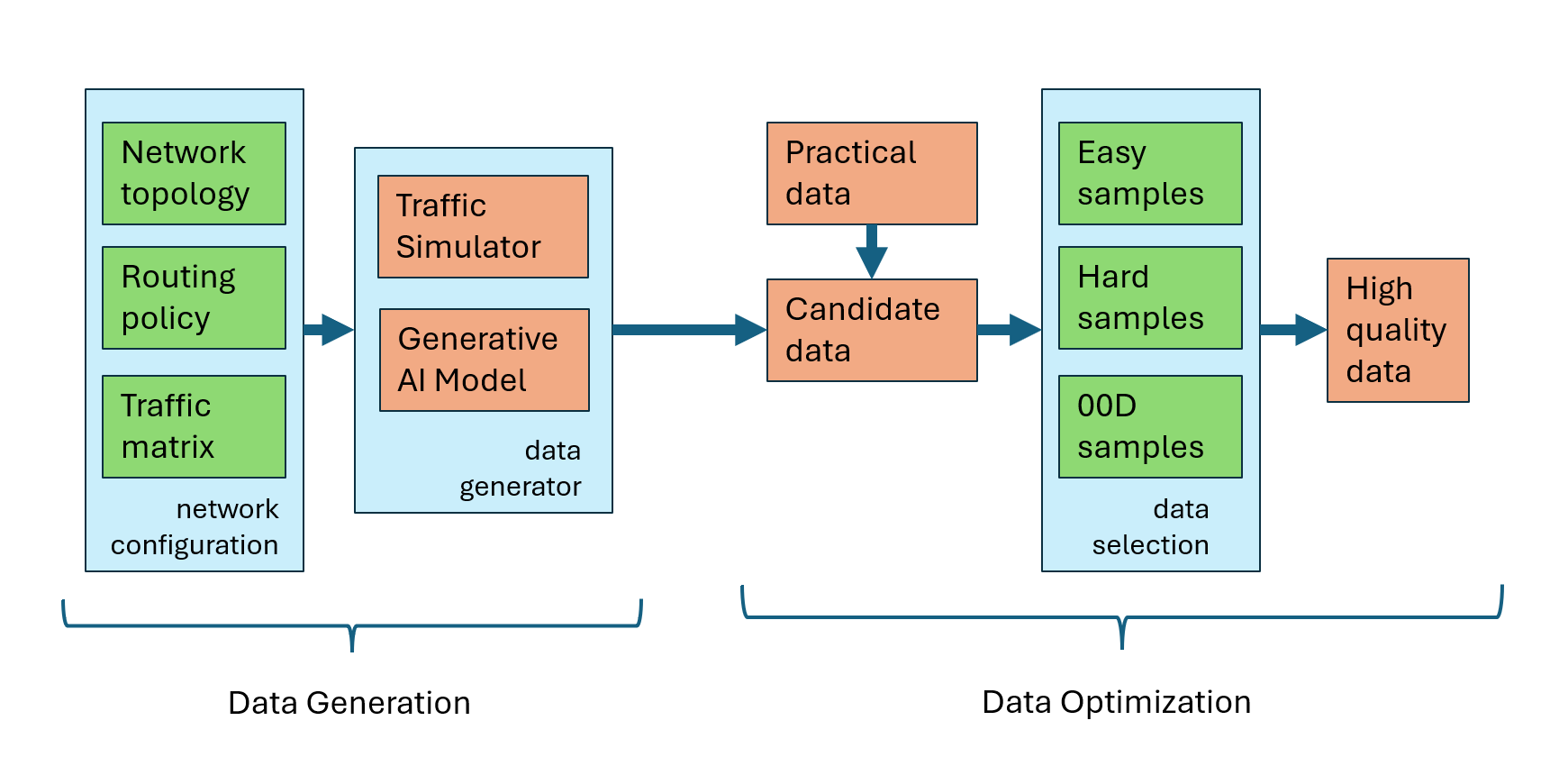}
    \caption{Framework of Data Generation and Optimization for NDT Performance Modeling (IETF \cite{draft-li-nmrg-dtn-data-generation-optimization})}
    \label{fig:enter-label}
\end{figure}

While the \gls{ndt} concept is elegant and straightforward, building and maintaining an \gls{ndt} on a real operational setup may involve significant investment in resources, infrastructure, skilled personnel, and ongoing calibration.

\subsection{Interaction between the Network Digital Twin and the Network Physical Twin}

One of the main challenges of a Network Twinning System is the calibration and compromise between model precision, data sources selection, data update frequency (granularity). Multiple data sources are usually available. The typical data available is Telemetry data that is collected through probes, agents etc. and may include metrics like traffic flow, latency, jitter, packet loss, and even environmental data from sensors. 

The data exchange between a \gls{ndt} and the physical network is a dynamic, two-way interaction. From one side it enables real-time synchronization, so that the \gls{ndt} is able to trustworthy represent the \gls{ndt} at each time, with an adequate degree of granularity and fidelity. Indeed, the virtual traffic simulated in the \gls{ndt} should be consistent with the real one, in terms of data path, performance KPIs, and data characteristics. This is a necessary condition to enable intelligent decision-making, prediction, what-if scenarios evaluation and monitoring. On the other side it this interaction is necessary to allow the \gls{ndt} to act on the physical network (directly or indirectly, e.g. through controller or orchestrator)  and enforce policies, for example for triggering correction or optimization actions, or directly sending reconfiguration commands or optimization suggestions. (This might involve rerouting traffic, adjusting resource allocation, or initiating maintenance actions.)

The data flows through the various levels of the Twinning System, starting from the data collection from the lower levels Physical Network, whether it's a mobile access network, transport network, or core infrastructure. The amount of available data can be quickly become overwhelming and impact on the network operation itself. The challenge is therefore to find the right compromise on the different aspects of data collection, aggregation, and transmission that suits the purpose of the twinning system.

The communication bridges that allow this interaction, and the exchange of control messages and data, are the Data Collection Interface and the Control Interfaces. A Network Twinning System should be designed for being first of all sustainable, to provide clear advantages to the management of the network.

\subsection{Data synchronization challenges}

Ideally, to the \gls{ndt} be kept in sync in real time with the \gls{npt}, a constant and timely flow of data is required. This means ultra-low latency and high-frequency data updates.

In practice, a compromise should be found, in order not to consume more resources than the ones that makes the system a sustainable one.  

To create a high-fidelity virtual replica of the network the \gls{ndt} uses the collected data (and historical data) in combination with a predictive model. This model is used to simulate the network behavior and predicting its future behavior, testing new configuration and analyzing its outcomes, without impacting the live production system. However, the predictive model is also fundamental for another important task, i.e. keeping the network state up to date between data updates.

The frequency of data updates is therefore linked with the precision of the model as well as how timely the system should react to external events (e.g. how likely external events can impact the KPIs of the network).  However, a high synchronization rate can overwhelm the network itself with data traffic, increasing the need of resources and creating traffic congestion. If we consider real work set ups, it should be taken into account that the network infrastructure has physical resource limitations, specially in mobile environments. These limit the bandwidth available and the capacity of processing the data, and introduce latency, thus possibly delaying updates and impacting the regular data transmission. Even in the case that a model is able compensate a high latency in the data updates, but the twinning system will still be slow to react to unexpected events, such as unpredictable traffic patterns or failures. 

On the other hand, when updates not frequent enough, the \gls{ndt} state can become outdated, reducing the fidelity of the twin and leading to poor predictions or wrong decisions.  This is specially true in scenarios where the networks topology changes in time like in vehicular networks or UAVs networks or 5G edge deployments. In such scenarios, if not properly configured, the twin’s capability to stay is sync cannot encompass the change rate, especially if the infrastructure isn’t optimized for high-frequency updates.

Some point out that "a digital twin is only as good as its model". While other state that ultra-low latency and high-frequency data updates are the key to ensuring that the \gls{ndt} accurately reflects the physical network’s behavior under all conditions. The main point is that we are considering a twinning system, and we should consider it as a whole when evaluating it goodness, which depends on the goal of the system, the conditions under which it is expected to operated, and the criticality (and involved risks) of the decisions that it takes.

\subsection{Scalabilty vs. Complexity}

Networks can become very complex systems, as the grow in size and diversity of technologies involved. In particular mobile networks have a high degree of complexity due to various reasons (radio access, mobility, diversity of \glspl{ue}, slicing, edge computing). Modellng nodes, links, and behavior of a network can become an computationally intensive task when the network grows, if not taken into account at the \gls{ndt} design time. To To allow the \gls{ndt} to scale accordingly, modeling strategies should be considered accurately.

\subsection{Security and Privacy Risks }

The \gls{ndt} has access to sensitive operational data. If not properly secured, it could become a target for cyberattacks or data breaches, especially when control commands are sent back to the physical network.

\subsection{Network Digital Twin Evaluation}

It is possible to leverage to metrics to understand the functioning of a \gls{ndt}, its characteristics and how well it is aligned with its physical counterpart. The \textit{Twin Alignment Ratio} has been proposed by to measure how well the digital twin mirrors the real system over time \cite{ckir2023synchronize}, and that is based on the ration of two numbers \textit{Planned Twinning Frequency}, corresponding to the frequency at which the twined device is configured to schedule the transmission of the data, and the \textit{Achieved Twinning Frequency}, corresponding to the frequency at which the digital twin received the packets sent from the twinned device. 

Consistency could directly affect the quality of the digital twining system. A consistency evaluation between the models and corresponding physical objects can be performed to indicate how closely the internal states (e.g. routing tables, buffer occupancy) of the \gls{ndt} match those of the physical network over time.  The \gls{sci} ensures that the \gls{ndt} mirrors the real-time state of network e.g. in terms of bandwidth allocation, latency thresholds, and user mobility patterns. If the SCI drops, it could mean a slice is misconfigured or drifting from its intended behavior. As an example, \cite{zhang2022consistency} provides a The framework of consistency evaluation for DTS models and discusses evaluation methods.

\gls{ndt} goodness can be measured using various types of indicators, in the following we indicate some of them. The \textit{Update Latency} takes into account the time that it takes for a change in the physical network to be reflected in the digital twin. Low latency means a better possibility to achieve a good synchronization. This is specially important for  \gls{urllc} use cases and mission-critical applications. Delayed or missed updates can compromise the capability of congestion prediction or of managing handover issues timely. The \textit{Data Freshness} (or Age of Information) express how the data received (or maintained by) the \gls{ndt} is recent and can be measured by using time-stamped updates. The \textit{Synchronization Frequency} indicates how often updates are exchanged: it can be fixed, e.g. at regular intervals, or it can change and adapt with network dynamics or \gls{ndt} changing requirements. Another indicator is the how much the behavior predicted by the \gls{ndt} deviates from the actual behavior in the physical network (\textit{deviation of prediction}).
\section{Designing a Network Digital Twin}

The key elements of the \gls{ndt} eco-system have been introduced in the IETF Internet Draft \cite{draft-irtf-nmrg-network-digital-twin-arch}. These are the following:
\begin{itemize}
  \item Data: Data is the most important element for building a digital twin. Data about the real network is collected and stored in a data repository. This include historical data and/or real-time data (configuration data, operational state data, topology data, trace data, metric data, process data, etc.) about the physical twin that are required by the models building and operation. 
  \item Models: Models are used to represent the behavior of the digital twin. There are two main types of models: base models and functional models. Base models represent the current state of the network, while functional models are used to simulate the behavior of the network under different conditions.
  \item Interfaces: Interfaces are used to communicate with the digital twin. Input interfaces allow users to input data and commands into the digital twin, while output interfaces provide information about the state and behavior of the digital twin.
  \item Mapping: Mapping is used to identify the digital twin and underlying entities and to establish a real-time interactive relationship between the real network and the twin network or between two twin networks.
  \item Logic to alayse, dignose, optimize, control and emulate.
\end{itemize}

The next sub-sections describe how to build the Network Digital Twin of a 5G private network, capable of precisely replicating the traffic from the actual physical infrastructure.

\subsection{The Physical Twin}
The Network Physical Twin (NPT) represents the actual infrastructure that will be replicated in the corresponding Network Digital Twin (NDT).
In this paper, the considered physical twin is the Amarisoft Callbox Classic \cite{amari186callbox-classic}. Such device allows the deployment of a 4G/5G private network. Amarisoft Callbox Classic consists of two built-in simcards which allow to connect two mobile 4G/5G cellphones, the \gls{ran} and the core network. All the hardware and software is built in the form of a single physical device. 

A single Amarisoft Classbox Classic is capable of supporting the configurations in Table~\ref{tab:amariconf}. More details, please refer to \cite{amari186callbox-classic}.
% \begin{itemize}
  % \item 4G LTE: 3 cells 20MHz 2x2, or 1 cell 20 MHz 4x4 + 1 cell 20MHz 2x2
  % \item 5G NR SA Mode: 1 5G cell 50MHz 2x2, or 3 cells 20MHz 2x2, or 3 cells 40 MHz SISO
  % \item 5G NR NSA Mode: 1 5G NR 50MHz 2x2 + 1 LTE 10MHz 2x2, or 1 cell 5G NR 40 MHz 2x2 + 1 cell LTE 20 MHz 2x2, or 1 cell 5G NR 40 MHz SISO + 2 cells LTE 20 MHz 2x2
% \end{itemize}

\begin{table}[H] 
%\small % Change table font size
\caption{Amarisoft Classbox Classic supported configurations.\label{tab:amariconf}}
%\isPreprints{\centering}{} % Only used for preprints
\begin{tabularx}{\textwidth}{XXXX}
\toprule
\textbf{Mode}	& \textbf{Configuration 1}	& \textbf{Configuration 2} 	& \textbf{Configuration 3}\\
\midrule
4G LTE & 3 cells 20MHz 2x2 & 1 cell 20 MHz 4x4 + 1 cell 20MHz 2x2 & \\ 
5G NR SA Mode & 1 5G cell 50MHz 2x2 & 3 cells 20MHz 2x2 & 3 cells 40 MHz SISO \\ 
5G NR NSA Mode & 1 5G NR 50MHz 2x2 + 1 LTE 10MHz 2x2 & 1 cell 5G NR 40 MHz 2x2 + 1 cell LTE 20 MHz 2x2 & 1 cell 5G NR 40 MHz SISO + 2 cells LTE 20 MHz 2x2 \\
\bottomrule
\end{tabularx}
\end{table}

\subsection{The Network Digital Twin technology}
The Network Digital Twin is built based on a network emulator, ComNetsEmu \cite{comnetsemu}. This software-based environment is built as a standalone virtual machine that combines an \gls{sdn} network emulator (Mininet \cite{mininet}) with an \gls{nfvi} solution (docker) into a docker-in-docker integrated framework.
In order to build the \gls{ndt}, the network emulator is used to deploy a containerized version of Open5Gs open source software for the 5G Core \cite{open5gs}  and UERANSIM for the 5G \gls{ran}  \cite{ueransim}.

\begin{figure}[t]
\includegraphics[width=12cm]{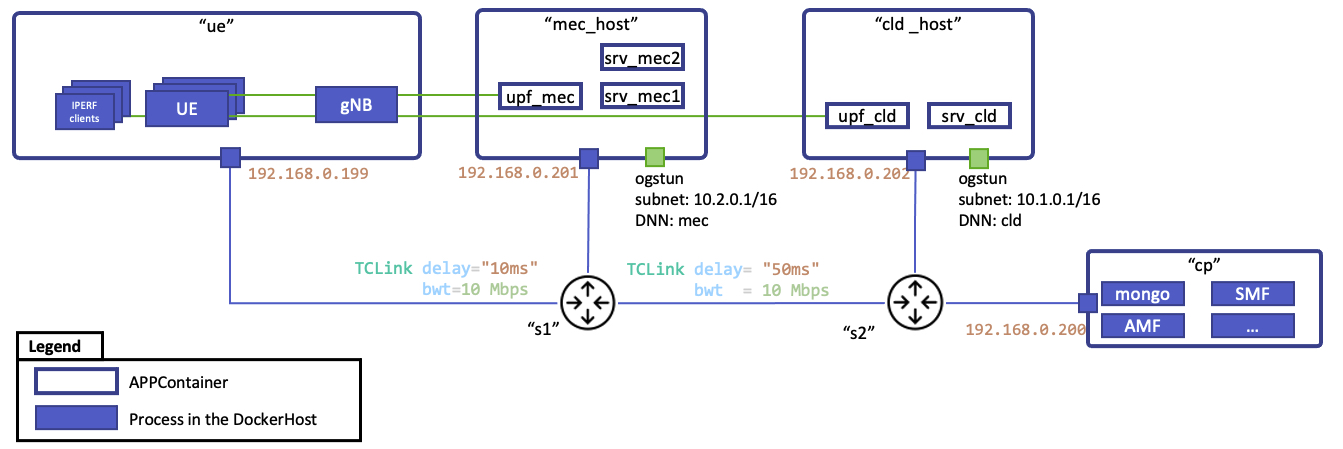}
\centering
\caption{Example of deployment of 5G open source implementations in ComNetsEmu.}
\label{fig:comnetsemu}
\end{figure}

In particular, Figure~\ref{fig:comnetsemu} shows the blueprint used for the deployment of the \gls{ndt}. Four virtual hosts (containers) are deployed, for emulating the 5G \gls{ran}, a \gls{mec} host, and two cloud hosts, respectively. The 5G \gls{ran}  docker container is used to deploy UERANSIM software, e.g. the UEs and \gls{gnb} that constitute the \gls{ran} . One cloud host is used for deploying services and the main \gls{upf} of the 5G \gls{sba}, while the other cloud host is used for the 5G Control Plane (NSSF, SMF, AMF, etc.). Finally, the \gls{mec} node is used to emulate \gls{mec} solutions, if needed.
The hosts are connected by virtual switches realized using OpenVSwitch software\footnote{\url{https://www.openvswitch.org}}, interconnected via 10Mbps links with variable latency. Both parameters are configurable by the configuration of the \gls{ndt} script.
Finally, the figure provides an example of two network slices (in green), one for \gls{mec} services and one for cloud services.

\begin{figure}[t]
\includegraphics[width=8cm]{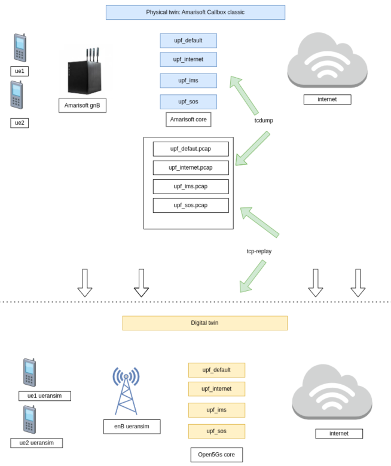}
\centering
\caption{Conceptual diagram of the connection between Physical and Digital Twin of the Amarisoft platform}
\label{fig:concept}
\end{figure}

\subsection{The Design Methodology and Approach to the Network Digital Twin}
The methodology to build and run the \gls{ndt} consists of three stages.

The first stage is data acquisition. This involves the collection of data from the \gls{npt}. The data collected is the configuration data to replicate the topology, user and status of the Physical Twin in the Digital Twin. Such configuration is then converted into configuration files for UERANSIM and Open5Gs software.

In the specific scenario considered in the paper, the configuration data can be retrieved by the \texttt{mme.cfg} configuration file of the Amarisoft device. Some of the parameters extracted include access point names, which represents the \gls{upf} names in the Open5Gs software, IP addresses, bandwidth of each tunnel, and the number of users. The data is extracted through a custom software, processed locally and save in a JSON file. The data is then trasmitted to the \gls{ndt} using the paramiko\footnote{Parimiko a pure-Python library implementing both client and server functionalities for the SSHv2 protocol, \url{https://www.paramiko.org}} library for SSH and SFTP communication. 

The following phase is modeling. The modeling phase is done in the ComNetsEmu network emulation environment. This is achieved by configuring and deploying all components of the 5G System by using proper Docker containers. Using the data provided in the first stage, the  slice configuration is replicated. In this phase, the slices described in JSON files are implemented. Such slices are characterized unique DNNs, IP addresses, subnets, bandwidths, and \gls{qci}. This information is written in the corresponding YAML files of Open5GS related to the configuration of the SMF, NSSF and AMF Network Functions of the 5G \gls{sba}. Such information is stored in the following files: \texttt{smf.yaml}, \texttt{nssf.yaml}, and \texttt{amf.yaml}. 

\subsection{Running the Network Digital Twin}
The last part to be configured is the \gls{ran}. The \gls{ran}  is represented by \glspl{gnb}, which manage radio communication with \glspl{ue}, handling tasks such as connection setup, mobility, and resource allocation. 

Once everything is set up, the network is started using the \texttt{net.start()} command.

The final stage is the synchronization stage. This stage represents the key feature of the \gls{ndt}, and it allows continuous communication between the Physical and the Digital Twin. 

To achieve this, the \gls{npt} (e.g. the Amarisoft Callbox Classic) is activated and \glspl{ue} are connected. Then, different traffic flows are generated using real applications and services, and captured by using the \textit{tcpdump} packet analyzer\footnote{\url{https://www.tcpdump.org}}. Captures are run continuously for \textit{T} seconds, and each capture file is stored in a dedicated folder in \textit{pcap} format on the Physical Twin device. The traffic is typically captured from the default network interface of the Physical Twin. It is usually called \textit{tun2}, but the name may be different depending on the content of the \texttt{mme.cfg} configuration file.

Meanwhile, the \gls{ndt} is activated, and it received the most recent \textit{pcap} traces from the folder on the Physical Twin via an SSH connection. The traffic is replayed on the \gls{ndt} by using the \textit{tcp-replay} software\footnote{\url{https://tcpreplay.appneta.com}}.

This means that the \gls{ndt} provides a picture of the real system with a delay equal to \textit{T} plus the communication delay between the Digital and the Physical Twin.
This delay can be mitigated, if necessary, by exploiting traffic prediction techniques, such as the ones analyzed in \cite{10765759}.

\section{Experimental Results}

Experiments were run in order to analyze the accuracy of the \gls{ndt} in replicating the behaviour of the \gls{npt} .

In the experiments, the \gls{ndt} and the \gls{npt} are connected via a 1Gbps Ethernet LAN. Tests are performed indoor.

The software used to implement the experiments in this paper and the detailed instructions are freely available through a GitHub repository\footnote{\url{https://github.com/fabrizio-granelli/Amarisoft.digital.twin}}.

Four scenarios with different traffic flows are considered in the experiments. Moreover, for sake of a better interpretation of the results, the clock of the \gls{ndt} is aligned with the clock of the \gls{npt}, so that the graphs can be easily compared. For the sake of consistency, in all the figures of this section we are going to maintain the following order: the graph on top shows the measured throughput on the \gls{npt}, and the one below shows the measurements on the \gls{ndt} for the same scenario. Finally, in all experiments, \textit{T}  = 120 seconds. 

\begin{figure}[t]
\includegraphics[width=8cm]{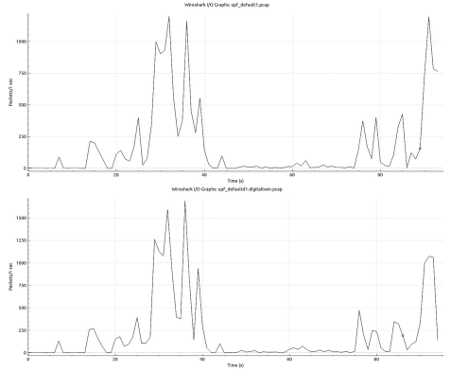}
\centering
\caption{Switching on the phones and browsing}
\label{fig:results-1}
\end{figure}

The first scenario consists of two \glspl{ue}, which are first turned on, registered to the network, and then used to browse (e.g. \url{google.com} website).
Figure~\ref{fig:results-1} shows the traffic captured from both the \gls{npt} and the \gls{ndt}. It is important to note that the two graphs appear extremely similar in the shape as well as in terms of time, thus demonstrating a good accuracy of the proposed solution in this scenario.

\begin{figure}[t]
\includegraphics[width=8cm]{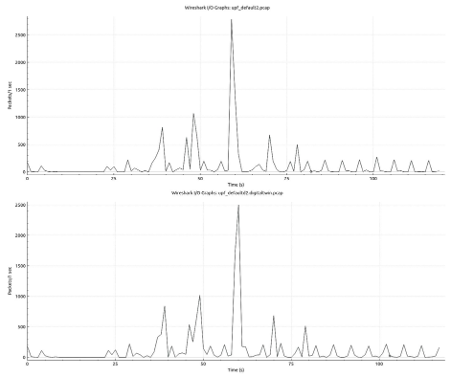}
\centering
\caption{Streaming from YouTube.com}
\label{fig:results-2}
\end{figure}

\begin{figure}[t]
\includegraphics[width=8cm]{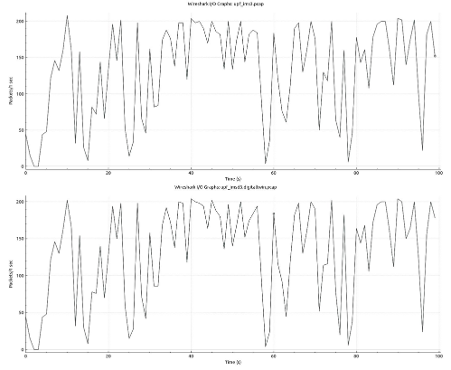}
\centering
\caption{Voice calls between 2 mobile phones}
\label{fig:results-3}
\end{figure}
 
\begin{figure}[t]
\includegraphics[width=8cm]{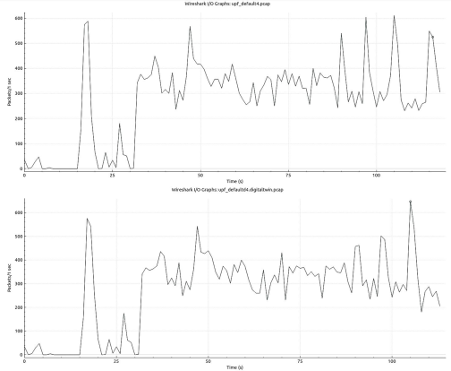}
\centering
\caption{Live broadcasting from facebook.com}
\label{fig:results-4}
\end{figure}

Figure~\ref{fig:results-2} shows the second scenario where the \glspl{ue} are streaming videos from youTube. Again, the resulting accuracy is clear visible by analyzing the figure.

Figure~\ref{fig:results-3} shows the third scenario. This time, voice calls are activated on the two \glspl{ue}. This means that in this scenario we are using a different interface on the Amarisoft Callbox Classic. This interface is the \gls{ims} interface which is again configured in the \texttt{mme.cfg} file. As clearly shown in the figure, the traffic captured was almost similar. 

In the last tested scenario, one \gls{ue} is used to  live stream a video on Facebook website. In this case, the generated traffic is mostly on the 5G uplink. As shown in Figure~\ref{fig:results-4}, it’s clear that the results are again accurate and extremely similar in both the \gls{npt} and the \gls{ndt}.

In conclusion, the experiments conducted on a real setup by using real 5G equipment demonstrate that the proposed strategy is capable of building an accurate \gls{ndt} of a 5G Private Network and to maintain effective synchronization between the \gls{npt} and its \gls{ndt}.

\section{Discussion, Open Challenges and Future Work}

The approach proposed in this paper has proven to provide a precise replica of what is happening in the physical twin in its digital counterpart. This allows to use the \gls{ndt}n to provide a precise picture of the behaviour and status of the actual physical network infrastructure.

In its current version, the \gls{ndt} can be used with data collection tools, such as Prometheus to capture and analyze different aspects of the network (e.g. monitoring performance or anomalies).

The use of \gls{ml} and \gls{ai} tools can also be implemented based on this project. Indeed, the \gls{ndt} can be used to generate endless accurate datasets that can be used to train \gls{ml} models, or used by learning algorithms in order to get a precise prediction of the impact of different management decisions BEFORE testing them on the actual network infrastructure.

However, several challenges and opportunities still exist in designing \gls{ndt}.

First of all, one point still open in the design of \gls{ndt} is which data should be transferred between the \gls{npt} and its digital replica. In this paper, we proposed to replicate the topology, setup and traffic from the physical infrastructure. However, this solution requires to transfer a relevant amount of data, that might introduce hard constraints on the capacity of the communication channel - as communication between the \gls{npt} and the \gls{ndt} might introduce additional delay in the replication of the traffic and thus represent a bottleneck for the performance of the \gls{ndt}.

Different type of data could be collected from the actual network infrastructure. For instance, the use of Prometheus to get metrics from the physical twin could be used to replicate its state in the digital twin. Another possibility could be to use network management protocols, such as \gls{snmp}, to collect data from managed devices and use such information to transfer the state between the physical and the digital entity.

To identify the nature of the data to be transferred is a problem still not properly addressed in the scientific literature. Indeed, the state-of-the-art lacks formal approaches to provide clear directions in this field. Proper methodologies will need to be defined in order to provide a framework for analyzing and comparing the different alternatives in order to understand the most suitable set of parameters to transfer and the frequency of updates, as well as their impact on the accuracy of the resulting \gls{ndt}. In this scenario, the application of information theory and semantic communications concepts will probably provide some interesting results.

Two other open points, that will be subject to further analysis by the authors, include:
\begin{itemize}
  \item \gls{ndt} state prediction: \gls{ml} or \gls{ai} can be used to enable to "\textit{look in the future}", i.e. by performing predictions about the state and traffic flows in the system, to provide a prediction about how the physical system might evolve in the (near) future. Nevertheless, the prediction capability can also be used to reduce the time delay between the clock in the \gls{ndt} and the actual clock on the physical infrastructure, by enabling to generate a "\textit{negative delay}" and thus being able to analyze the future of the physical twin. Please see \cite{10978133} for a comprehensive discussion about reliable prediction in \glspl{ndt}.
  \item Implementation of the bidirectional connection: in this work, an one-directional link from the physical infrastructure to the \gls{ndt} is considered. However, the link in the opposite direction, i.e. from the \gls{ndt} to the physical infrastructure, can be implemented. In this case, the most common application would be to replicate changes successfully introduced in the \gls{ndt} on the actual network infrastructure. This will open new opportunities to control the physical system and to "close" the control loop typical of several network automation approaches.
\end{itemize}

\section{Conclusions}

This paper discusses the issues in building an effective \gls{ndt} for 5G systems, focusing on a test case related to a 5G Private Network. The authors propose a methodology that allows to extract topology and configuration information in order to replicate the network setup from the actual physical network infrastructure to the \gls{ndt}. The \gls{ndt}n is built based on a network emulator and docker containers. Traffic flows are replicated in order to maintain synchronization between the Physical and the Digital Twin.
Experiments performed in the lab by using an actual indoor 5G private network deployment demonstrate high accuracy in replicating the behaviour of the Physical infrastructure into its \gls{ndt}.
Future work will be aimed at reducing the time shift between the Physical and the Digital Twin, as well as characterizing the bidirectional link between the two entities, in terms of both performance requirements and functionalities.

\vspace{6pt}

\section{Acknowledgements}
This research was funded by the European Union under NextGenerationEU PRIN 2022 6GTWINS project (Prot. n. 2022MWBFEE), by the European Union’s Horizon Europe under Grant Agreement no. 101096342 (HORSE project),  Grant Agreement No 101191436 (MARE project), and Grant Agreement No 101192080 (6G-LEADER project), and by the European Union under the Italian National Recovery and Resilience Plan (NRRP) of NextGenerationEU, partnership on “Telecommunications of the Future\ (PE00000001 - program “RESTART”).

\section{Software Availability}
The software developed in this project is available as open source at \url{https://github.com/fabrizio-granelli/Amarisoft.digital.twin}.

%\section{Abbreviations}

%The following abbreviations are used in this manuscript:
%\\
%\printglossary[type=\acronymtype,nonumberlist]

\bibliographystyle{unsrt}  
\bibliography{references}  %%% Remove comment to use the external .bib file (using bibtex).

%\printbibliography

\end{document}